\documentclass[aps,manuscript,prl,english,reprint,onecolumn]{revtex4-1}
\usepackage[T1]{fontenc}
\usepackage[latin9]{inputenc}
\usepackage{amsmath}
\usepackage{amssymb}
\usepackage{array}
\usepackage{graphicx}
\usepackage{verbatim}
\usepackage{floatrow}
\usepackage{booktabs}
\usepackage{multirow}
\usepackage{hyperref}
\usepackage{scrextend}
\usepackage{verbatim}
\usepackage{comment}
\usepackage{blindtext}
\usepackage[skins,theorems,most]{tcolorbox}
\makeatletter
\usepackage[table]{xcolor}

\usepackage{babel}

\newcolumntype{P}[1]{>{\centering\arraybackslash}p{#1}}

\DeclareMathAlphabet{\mathpzc}{OT1}{pzc}{m}{it}
\renewcommand*\l@subsection{\@dottedtocline{2}{1.8em}{3.2 em}}
\renewcommand*\l@subsubsection{\@dottedtocline{2}{1.8em}{6.4 em}}

\g@addto@macro\normalsize{%
  \setlength\abovedisplayskip{0pt}
  \setlength\belowdisplayskip{0pt}
  \setlength\abovedisplayshortskip{0pt}
 \setlength\belowdisplayshortskip{0pt}
}

\usepackage[T1]{fontenc}
\usepackage[latin9]{inputenc}
\usepackage[autopunct=true]{csquotes}
\usepackage{amsmath}
\usepackage{amssymb}
\usepackage{array}
\usepackage{graphicx}
\usepackage{esvect}
\usepackage{verbatim}
\usepackage{floatrow}
\usepackage{booktabs}
\usepackage{multirow}
\usepackage{hyperref}
\usepackage{scrextend}
\usepackage{blindtext}
\usepackage{amssymb}
\usepackage{amsfonts}
\usepackage{amssymb}

\makeatletter
\usepackage{babel}

\newcolumntype{P}[1]{>{\centering\arraybackslash}p{#1}}

\DeclareMathAlphabet{\mathpzc}{OT1}{pzc}{m}{it}
\renewcommand*\l@subsection{\@dottedtocline{2}{1.8em}{3.2 em}}
\renewcommand*\l@subsubsection{\@dottedtocline{2}{1.8em}{6.4 em}}

\g@addto@macro\normalsize{%
  \setlength\abovedisplayskip{0pt}
  \setlength\belowdisplayskip{0pt}
  \setlength\abovedisplayshortskip{0pt}
 \setlength\belowdisplayshortskip{0pt}
}

\tcbset{
  myyellowbox/.style={
    colback=yellow!10,    
    colframe=yellow!50!black, 
    boxrule=0.5pt,
    arc=2pt,
    left=6pt,
    right=6pt,
    top=4pt,
    bottom=4pt
  }
}

\usepackage{graphicx}
\usepackage{float}

\usepackage{derivative}
\usepackage{xcolor}
\usepackage[margin=0.68in]{geometry}
\usepackage{siunitx}
\usepackage{soul}
\usepackage{minitoc}
\usepackage{color}
\usepackage{amssymb}
\usepackage{amsmath, scalerel, breqn}
\usepackage{physics}
\usepackage{graphicx}

\definecolor{dark-red}{rgb}{0.6,0.1,0.1}
\definecolor{dark-blue}{rgb}{0.15,0.15,0.8}
\definecolor{RED}{rgb}{0.55,0.1,0.1}
\definecolor{GREEN}{rgb}{0.0,0.75,0.0}
\definecolor{BLUE}{rgb}{0,0,0.75}
\definecolor{medium-blue}{rgb}{0,0,0.5}
\usepackage{hyperref}
\hypersetup{
    colorlinks, linkcolor={dark-red},
    citecolor={dark-blue}, urlcolor={medium-blue}
}

\begin{document}

\begin{titlepage}

    \begin{center}
        {\large {Discovering novel quantum dynamics with NISQ simulators } \par}
        \vspace{2mm}
        {\small
        P. Roushan\,(pedramr@google.com) and L. Martin\,(leighmartin@google.com)
        
        \vspace{1.mm}
      Google Quantum AI, Santa Barbara, California, USA
      
      \vspace{1.mm} \today }
   \end{center}
   \global\let\newpagegood\newpage
  \global\let\newpage\relax
\end{titlepage}

\global\let\newpage\newpagegood

\vspace{12mm}


Major technological advances of the past century are rooted in our understanding of quantum physics in the non-interacting limit. A central challenge today is to understand the behavior of complex quantum many-body systems, where interactions play an essential role.  About four decades ago, Richard Feynman proposed using controllable quantum systems to efficiently simulate complex physics and chemistry problems, envisioning quantum orreries, highly tunable quantum devices built to emulate less understood quantum systems \cite{feynman1982simulating}. Here we ask whether quantum simulators have already uncovered new physical phenomena-and, if so, in which areas and with what impact. We find that, in several notable instances, they have advanced our understanding of many-body quantum dynamics. Although many of these insights could in principle have been obtained theoretically or numerically, they were nevertheless first achieved using quantum processors. While a broad landscape of problems beyond non-equilibrium dynamics still awaits exploration, it is encouraging that quantum simulators are already beginning to challenge and refine our conventional wisdom.

\vspace{2mm}
Modern quantum processors occupy an intermediate regime: some offer high-fidelity control over tens of qubits, while others provide larger qubit counts at the expense of coherence. This landscape defines the Noisy Intermediate-Scale Quantum (NISQ) era \cite{preskill2018quantum}, where individual qubit control is routine, yet decoherence constrains circuit complexity. Despite these limitations, the past decade has brought substantial advances in manipulating large, interacting quantum many-body systems.

\vspace{10mm}

\begin{figure*}[h!!]
\centering
\includegraphics[width=0.75\textwidth]{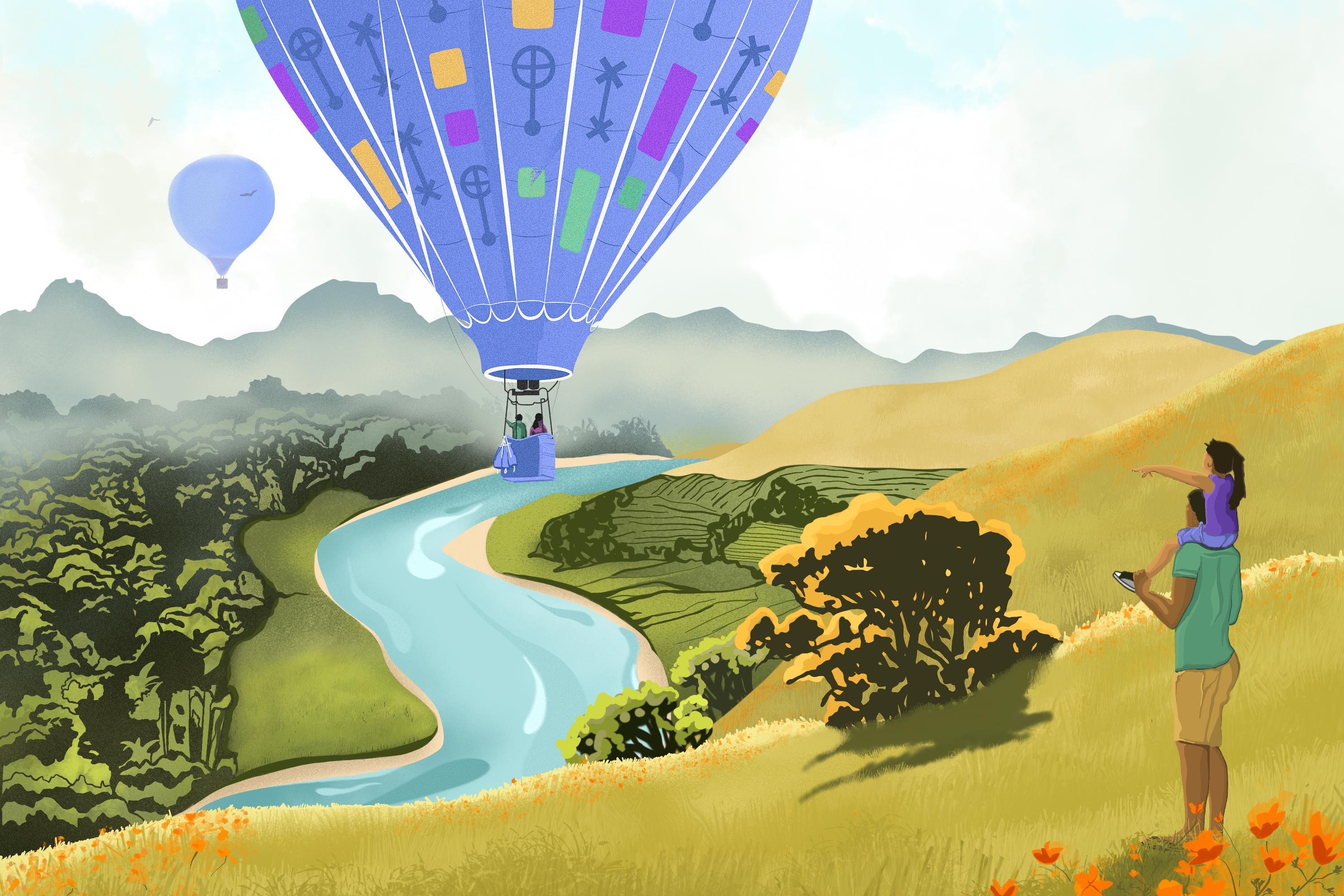}
\end{figure*}

Current quantum processors are somewhat like hot-air balloons: they require considerable effort to launch, and their maneuverability is limited, yet once aloft they can drift into regions that have never been explored before. When they do, they sometimes uncover small but genuine discoveries-discoverinos. Often, after such a region is identified, theoretical or numerical methods can later chart a more direct path to it. But the key point remains: it was the balloon that first touched down on the new terrain.

\newpage
At the dawn of the last century, quantum mechanics, the most accurate and successful theoretical framework for describing natural phenomena, was born. ``Normal'' science progress within this theory\,\cite{Kuhn62} enables us to achieve a comprehensive understanding of much of the physical world, from atoms and solid-state materials to elementary particles. This understanding led to the so-called ``first quantum revolution'', which profoundly transformed our lives through the invention of semiconductor and laser-based devices\,\cite{Milburn,Deutsch2020}. Nevertheless, most of what we have learned and utilized has resulted from major simplification of complex systems into systems of independent individual particles. For example, when a single electron propagates through a crystal, we approximate its interaction with other electrons only via an average background field. With this approximation we can adequately explain much of the conductance seen in metals and semiconductors. 

\vspace{1mm}
There are many physical phenomena that cannot be accurately described within a single-particle approximation; the inherently many-body nature of the problem must be taken into account. This is the underlying physics for large molecules and collective behavior observed in condensed matter settings, e. g.~ correlated electrons. In these cases, writing microscopic equations that take interactions into account is rather trivial, but solving them is extremely challenging. This was realized as early as 1929 by P. Dirac. In the modern technical parlance, when the constituent elements of a system become highly entangled, they exhibit extraordinary complexity that cannot be efficiently simulated even with the most powerful supercomputers or comprehended with our current theories. This ``entanglement explosion'' regime requires considering exponentially large configurations\,(\,Hilbert space dimension\,), which poses a challenge to classical computers.

\vspace{1mm}
In early 1908's, Feynman envisioned a solution for the challenge of simulating complex physics and chemistry problems\,\cite{Feynman82}. He proposed building quantum orreries: using highly controllable quantum devices to simulate less understood quantum systems. In 1996, Seth Lloyd showed that\,\cite{Lloyd96},  when the interactions among particles are local, the time needed for the simulation would grow polynomially --and not exponentially--with the number of particles. Some quantum simulation proposals consider a stroboscopic time evolution, i.e. digital, and some consider an analog evolution. Currently, researchers are investing heavily in both approaches. In almost parallel effort, most of the mathematical foundation for quantum algorithms and error correction was laid out in 1990s. This rigorous theoretical framework sparked research investments that spread beyond academic labs and led to large-scale industrial engagements and startup companies. 

\vspace{1mm}
A fault-tolerant quantum computer holds the promise of solving many scientific problems for which there are quantum algorithms, ranging from combinatorial optimization to nonlinear differential equations. Highly controlled and coherent quantum simulators offer similar promise without the demanding requirements of fault-tolerance. However, both of these technologies currently stand as future goals rather than present realities. Instead, we currently have moderately-sized NISQ processors, in which control of individual degrees of freedom (superconducting qubits, atoms, crystalline defects, ...), is possible, but decoherence limits the scale and complexity of device operation. 

\vspace{1mm}
In the last decade, we have witnessed major advances in quantum control on various NISQ platforms. This progress has led to exploration of many foundational physical effects, and has even played a role in the elucidation of recently discovered phenomena, particularly in the field of non-equilibrium quantum phases of matter. Key examples include the demonstration of many-body localization,\cite{AbaninRMP2019} and time crystalline behavior\cite{khemani2019brief}. For developers of quantum hardware, such experiments serve to validate the results of their devices, identify key limitations, and chart the path ahead for improved fidelity and desired capabilities.

\vspace{1mm}
While NISQ demonstrations play an essential role in the field's progress, they often reproduce rather than challenge conventional wisdom. In the last two decades, we have witnessed major advances in quantum control on various NISQ platforms. As a result, many \textit{a posteriori}, \textit{i.e.}, ``post-theory'' experiments have enabled us to identify limitations and capabilities of the platforms; examples include demonstration of many-body localization (MBL)\,\cite{AbaninRMP2019} and time crystalline behaviour\,\cite{khemani2019brief} as non-equilibrium quantum phases of matter. These experiments often reproduce conventional wisdom in progressively more complex problems, rather than challenging it.
However, we have noticed that a few experiments on NISQ processors have discovered unanticipated subtleties in non-equilibrium and near equilibrium quantum dynamics. Although decoherence errors and relatively small system sizes have thus far limited most such experiments to within the simulation capabilities of a classical computer, it nevertheless stands that several of these observations were first made on quantum processors \textit{a priori}-that is before they were identified theoretically. These occurrences offer a preview of what we stand to learn, and how the nature of experimental work in quantum information may change as hardware improves.

\newpage
Here, we mention a few works that exemplify and clarify our claim of NISQ-enabled discoveries, which we define as findings that were unknown to those working in a given field, and have significant impact on their discipline. Given the experiments up to now, we see NISQ discoveries as falling into three subjective categories, perhaps best illustrated by example:

\vspace{1mm}
\textbf{(i) Challenging conventional wisdom.} The first notable example of a discovery is the observation of ``quantum many-body scars'' in a one-dimensional\,(1D) chain of atoms\,\cite{bernien2017probing}. The system is first initialized in a spatially alternating pattern of low and high energy states\,(\,Fig.\,1\,(A)\,). Standard thermodynamics suggests that systems of sufficient complexity should relax monotonically to thermal equilibrium, but instead it was observed that the atoms continually entangled and disentangled, with the initial pattern disappearing and reappearing periodically. It was further observed, counterintuitively, that the these oscillations could be stabilized via periodic driving\,\cite{bluvstein2021controlling}. This reviving pattern, termed quantum many-body scarring, and has been debated in many subsequent works. Proposed explanations attribute its existence to closeness to integrability, or to the existence of an algebraic structure known as dynamical symmetries. Despite this progress, a general understanding of when a non-integrable quantum system can be expected to host many-body scars remains an open question. Quantum scars could suggest a new universality class beyond known non-equilibrium phases of matter such as many-body localized systems and conventional ergodic or integrable systems. 

\vspace{1mm}
Another notable example is the observation of resilient bound states to integrability-breaking perturbations, using superconducting circuits. Recently, it was found theoretically that an integrable model of quantum dynamics has stable quasiparticles and can form bound states\,\cite{Prosen2019}. Morvan\,\textit{et al.}\,\cite{morvan2022formation} used a ring of qubits to simulate these bound states with microwave photons\,(\,Fig.\,1\,(B)\,) . The resulting photon bound states are expected to be stable in integrable systems but to decay quickly when perturbed. However, when breaking integrability by introducing a quasi-1D geometry, they found that the decay of these bound states was unexpectedly slow. This surprising finding has invited further analytical and numerical studies of the problem\,\cite{Surace_PRX_2024,Papic_PRX_2024}. While the numerical studies confirm the quantum simulation's findings, a simple agreed-upon explanation is still lacking.

\vspace{1mm}
The last example we mention in this category is the unexpected propagation of quantum correlations in spin chains with tunable interaction range, simulated in a linear array of trapped $^{171}$Yb ions\,\cite{Richerme,jurcevic2014quasiparticle}. They observed faster-than-linear light-cone growth for the relatively short-range interaction\,(\,Fig.\,1\,(C)\,). Although faster-than-linear growth is expected for relatively longer-range interaction, there is currently no consensus as to whether such behavior is generically expected in the parameter regime that it was observed. 

\vspace{1mm}
\textbf{(ii) Testing scientific conjectures.} In several cases, quantum simulators have taken part an active role in a debated research question. The instances that we have identified in this regard are related to universal aspects of dynamics in 1D quantum magnets. For a wide range of classical systems, the growth of interfaces can be described by the Kardar-Parisi-Zhang (KPZ) equation, and belongs to the KPZ universality 
class. KPZ universality was also conjectured to describe spin transport in the 1D Heisenberg model based on theoretical evidence\,\cite{Ljubotina_PRL_2019}. This conjecture has been examined in two works that study the probability distribution of magnetic domain-wall relaxation across a 1D chain's center. Using a chain of $^{87}$Rb atoms, Wei \textit{et al.}\,\cite{wei2022quantum} found that up to the third (skewness) moment of transferred magnetization are indeed governed by the KPZ dynamical exponent\,(filled circles in \,Fig.\,1\,(D)\,). Using a chain of superconducting qubits, Rosenberg \textit{et al.} confirm these findings\,\cite{Eliott2024}. However, they find that when they start with initial states that are closer to equilibrium, skewness deviates from the predictions of the KPZ conjecture\,(square markers). They also examined the fourth moments\,(kurtosis) and found it does not fall under the KPZ universality on the timescales of their experiment. These results suggest that our current theoretical picture is inadequate to describe the dynamics in Heisenberg spin chains.  

\vspace{1mm}
\textbf{(iii) Computing physical parameters in correlated electron phases.}
These quantum simulation examples realize the Fermi-Hubbard model, a key model in studying correlated electron systems, whose phases are among the longest-standing unsolved problems in condensed matter. In high-temperature superconductors, the nature of charge and spin propagation through strongly interacting solids is poorly understood and difficult to probe. Nichols \textit{et al.}\,\cite{nichols2019spin} measured the spin diffusion coefficient and spin conductance in the Fermi-Hubbard model experimentally, not in a solid state system but rather in a two-dimensional optical lattice of fermionic atoms\,Fig.\,1\,(E)\,). Brown \textit{et al.} \,\cite{brown2019bad} measured the dependence of the resistivity on temperature, which shows a linear relation also encountered in strange metals\,(Fig.\,1\,(F)\,). Their data probe resistivity and diffusion constants in a parameter regime that challenges current numerical methods, and thus explore transport in a regime that had not been predicted theoretically. It is hoped that such experiments will eventually be able to probe microscopic dynamics in the Fermi-Hubbard model's hypothesized superconducting phase, thus helping to elucidate the underlying mechanism for high-temperature superconductivity.

\newpage
\textbf{Outlook.} As control over larger numbers of coherent qubits improves, we are witnessing experiments that were not possible even a few years ago. The upward spiral of hardware progress is slowly but steadily closing the gap between hardware capabilities and NISQ-friendly algorithms. While it is rather difficult to trace the history to make a one-to-one correspondence between technical progress and discovery instances, we can at least observe that today's discoveries are the result of many years of hardware advances across different platforms.

\vspace{1mm}
A main challenge in these experiments is imposed by decoherence in NISQ devices, which buries the signal in noise and makes it harder to extract observables. Some of the notable developments that have played a crucial role in advancing NISQ experiments are randomized measurement of entanglement entropy,\cite{ZollerScience2019}, as well as various error-mitigation techniques, such as those implemented in reference \,\cite{Kandala_nature_2019}. Ultimately, if recent progress in qubit coherence and operation fidelity continues, such decoherence-induced noise and control errors will diminish to the point where NISQ simulators can unambiguously exceed the capabilities of classical computers.

As quantum hardware becomes more capable, there will be opportunities to conduct research of a more exploratory nature, in which experiments are undertaken without a prior theoretical expectation for what precise phenomena will be observed. These experiments require reliable hardware and a careful understanding of its decoherence and other limitations, as the burden of proof falls on the experimentalist to show that any unexpected dynamics are not mere artifact. We anticipate this exploratory frontier will bring the problem of verification to focus, a concept which so far has been mainly discussed in the computer science parlance\,\cite{Gheorghiu2019Verification}. Experimental work beyond the limits of modern theoretical tools will pose significant challenges, but these challenges also offers the possibility for true scientific discovery.

\vspace{1mm}
Fault-tolerant quantum computation and fully controlled large-scale quantum simulation are among the major technological challenges of our time. These goals have inspired and engaged a large community of researchers over decades, placing these efforts among the biggest contemporary science projects of our time, and spreading them across academic, national, and industrial labs. It is invigorating to see that our collective effort thus far has born some fruits of discovery. The experiments thus far are heavily guided by theoretical results and state-of-the-art numerics, but they have begun to contribute alongside these methods as well. Leveraging their exponential computational gain with size, we soon enter a new phase in which NISQ processors will lead the path more directly and independently. Full-fledged discoveries and applications are ahead and awaiting us.

\begin{figure*}[t!]
\centering
\includegraphics[width=0.96\textwidth]{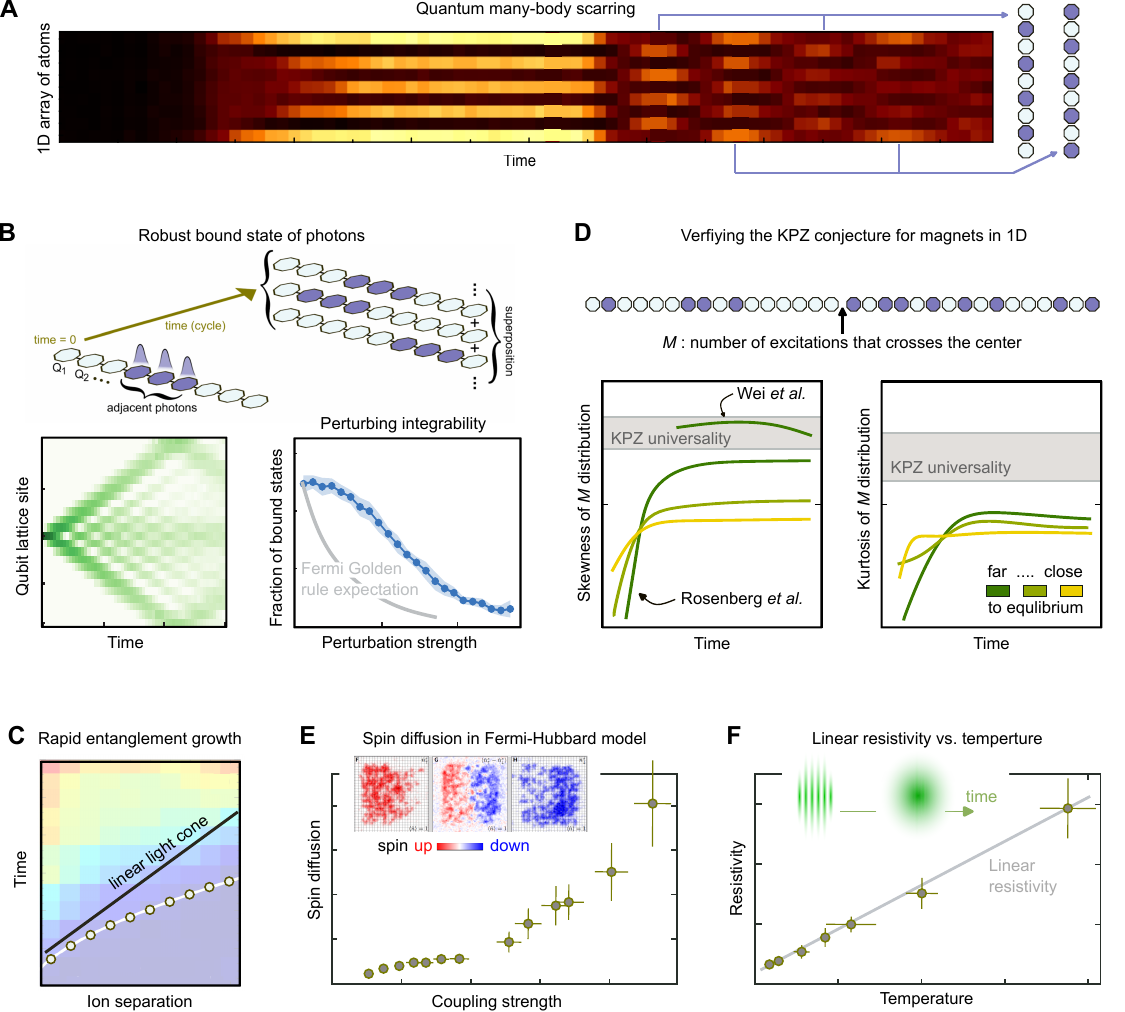}
\caption{ \textbf{Instances of NISQ discoveries.}\,\textbf{(A)} A 1D chain of 9 Rb atoms is initialized in a spatially alternating pattern of low and high energy Rydberg states\,\cite{bernien2017probing}. Single-atom excitation probabilities show an oscillatory pattern of entanglement and disentanglement, rather than monotonically equilibrating. \,\textbf{(B)} In a ring of 24 superconducting qubits and for a certain parameter range, the system is integrable, and initially adjacent excitations (microwave photons) stay together at all times\,\cite{morvan2022formation} (left, green interference pattern). Introducing an integrabilty-breaking perturbation is anticipated to break these bound states; however, even for a sizable perturbations the photons stay together, in contrast to a simple application of Fermi-golden rule.\,\textbf{(C)} Two-point correlation measurement in a chain of 11 $^{171}$Yb ions\,\cite{Richerme}, which simulates long-range XY model, shows faster-than-linear\,(white line) growth of the light-cone boundary.\,\textbf{(D)} Using a a chain of 50 $^{87}$Rb atoms, Wei \textit{et al.}\,\cite{wei2022quantum} measured the spin states of a Heisenberg chain (top) by removing one spin species\,(center) and imaged the atomic site occupation\,(bottom), and computed the probability distribution of magnetic domain-wall relaxation across a 1D chain's center. The first, second (not shown) and third moment (filled circle markers) of transferred magnetization are in the KPZ universality class. Using a chain of 46 superconducting qubits (square markers), it was found that skewness and kurtosis (fourth moments) increasingly tend toward zero vs. time as the initial states are taken closer to infinite temperature\,\cite{Eliott2024}.\,\textbf{(E)} Direct measurement of spin diffusion coefficient of the half-filled Fermi-Hubbard system\,\cite{nichols2019spin}. \,\textbf{(F)} Fluorescence image of the average density of spin polarization is used to extract resistivity data\,\cite{brown2019bad}. Initially the system is in thermal equilibrium with a spatially modulated density. Immediately after a sinusoidal potential is turned off, the system is no longer in equilibrium but the density has not yet changed. The rate of relaxation toward equilibrium enables extraction of resistivity.}
\end{figure*}

\bibliography{References.bib}
\end{document}